\documentclass[a4paper,10pt,twocolumn]{article}
\usepackage[utf8]{inputenc}
\usepackage{graphicx}
\usepackage{titling}
\usepackage[margin=0.85in]{geometry}
\newcommand{\subtitle}[1]{%
  \posttitle{%
    \par\end{center}
    \begin{center}\large#1\end{center}
    \vskip0.5em}%
}

\title{Dynamic Partitioning of Physical Memory Among Virtual Machines}
\subtitle{ASMI:Architectural Support for Memory Isolation}
\author{
  Jithin R\\
  \texttt{jithinr550@gmail.com}\\
  National Institute of Technology Calicut, India
  \and
  Priya Chandran\\
  \texttt{priya@nitc.ac.in}\\
  National Institute of Technology Calicut, India
}

\date{}
\begin{document}

\maketitle

\begin{abstract}
Cloud computing relies on secure and efficient virtualization. Software level security solutions compromise the performance of virtual machines (VMs), as a large amount of computational power would be utilized for running the security modules. Moreover, software solutions are only as secure as the level that they work on. For example a security module on a hypervisor cannot provide security in the presence of an infected hypervisor. It is a  challenge for  virtualization technology architects to enhance the security of VMs without degrading their performance. Currently available server machines are not fully equipped to support a secure VM environment without compromising on performance. A few hardware modifications have been introduced by manufactures like Intel and AMD to provide a secure VM environment with low performance degradation. In this paper we propose a novel memory architecture model named \textit{ Architectural Support for Memory Isolation(ASMI)}, that can achieve a true isolated physical 
memory region to each VM without degrading  performance. Along with true memory isolation, ASMI is designed to provide  lower memory access times, better utilization of available memory, support for DMA isolation and  support for platform independence for users of VMs.
\end{abstract}

\section{Introduction}
 In a  server system, sharing resources among virtual machines (VMs) implies that the resources of the server are accessed simultaneously among different users. Users' concerns about physical device sharing, stem from the fact that their data resides in these shared devices, and virtualization technology has to provide the necessary mechanisms to ensure the  security of user data. Protection of user data at different levels of architecture like CPU, memory and Input/Output (I/O) devices has to be provided, proved  and assured to convince the users of the credibility of the system. The survey reports \cite{jithin2014virtual} \cite{pearce2013virtualization} \cite{rehman2013virtual} on the security of VMs,  show that 
\begin{itemize}
  \item VMs  should be as isolated as physical machines, i.e., the only means of communication between them should be exactly as between two different physical machines (Isolation Property of VMs)
 \item Hypervisors cannot be trusted, due to the possibility of hypervisor based malware, and other infections possible on hypervisors
\end{itemize}
 An isolated VM environment can be provided by the hypervisor, either through hardware modifications or through additional software modules that restrain a VM from accessing  other VMs' data at various architectural levels \cite{jithin2014virtual}. Software modules to impose these restrictions would take additional CPU clock cycles and thereby  reduce  VM performance, implying that software solutions could  provide security to VMs only at the cost of compromising  VM performance. A hardware modification could impose these restraints with a lesser performance degradation, as it would use either a marginal number of additional clock cycles, or simply a negligible increase in the pipeline cycle time. Hence, we aim at designing enhancements to  hardware that can achieve security for  VMs without compromising on their performance, and also provide security in the presence of compromised hypervisors.

 This paper reviews literature in the area of memory virtualization to identify  open challenges that prevent hypervisors from providing a true isolated environment for VMs without losing out on performance. Some open challenges have been identified and  a  memory architecture model aimed at  solving  those challenges is proposed, which unlike existing memory models like nested paging and IOMMU, does not degrade performance. ASMI (Architectural Support for Memory Isolation),  our proposed  model, requires modification to hardware and is illustrated with the help of hardware currently in  popular use for supporting VMs.

 Memory virtualization techniques used in present day systems have been reviewed in the Section \ref{memory}, along with the open challenges in memory virtualization. Necessary features in any technology designed  for improving VMs in terms of security and performance are identified in Section \ref{solution}, along with a description of the design of ASMI. Section  \ref{conclusion} concludes the article with future directions of research.

  \section{Memory Virtualization} \label{memory}

  The normal paging mechanism cannot be used \cite{Smith2006} in a virtualized environment. In a normal computing environment, the paging hardware is accessed by the single operating system to get the physical address corresponding to the virtual address. In a virtual environment, a single physical memory is shared among different VMs or guest operating systems (hereafter referred as guest OS, in this paper). 
  
  If a guest OS accesses the paging hardware to translate its virtual address, there are possibilities that a guest OS may access a physical address which is previously accessed by another guest OS. As this is a serious security threat to the isolation property of VMs, the security of VMs at memory level, has to be assured by isolating the physical pages of each VM from others. A mechanism  to differentiate the physical address of each guest OS is a mandatory requirement.

  A memory virtualization mechanism that is popularly used in VM technology nowadays to achieve differentiation of logical addresses is called \textit{Nested Paging \cite{Smith2006}}. The following section describes nested paging.

  \subsection{Nested Paging}

  In nested paging \cite{Smith2006}, the virtual address of the process running in each VM is converted to a \textit{pseudo physical address} by the corresponding guest OS. This translation is achieved by looking up   \emph{page tables}, which are maintained by the guest OS \cite{Smith2006}. In the next level of translation, the pseudo physical address stored in page tables of each guest OS is translated to the actual physical address by the hypervisor. This translation is stored in the \textit{real map table} maintained by the hypervisor \cite{Smith2006}. The hypervisor maintains  a separate real map table for each guest OS. These two page tables, in combination, are known as a \textit{nested page table}. Thus, in a virtual environment, there are three layers of memory \cite{Smith2006}. They are \emph{Physical Memory} (Original device memory, visible to hypervisor), \emph{Pseudo Physical Memory} (Virtual view of physical memory to VMs, visible to guest OS) and \emph{Virtual Memory} (Logical memory for each 
program, visible to process)

  On contemporary platforms, page translation is supported by a combination of page table and a translation lookaside buffer (TLB). Currently there are two different types of hardware configurations available for address translation. One is  an architectured page table. The second is an architectured TLB \cite{Smith2006}. Virtualizing these hardware configurations is a primary requirement for memory virtualization.

  Each guest OS maintains its own page tables. These page tables represent the virtual to pseudo physical mapping that guest OS manages. To virtualize the architectured page table, a \textit{shadow page table} \cite{Smith2006} is used by the hypervisor, which contains the virtual address and the corresponding physical address. Each VM should have a shadow page table. When a context switch (among VMs) occurs, the shadow page table in use by the hypervisor changes according to the currently active VM \cite{Smith2006}. Shadow page tables are updated along with the nested page tables.

  In an architectured TLB, TLB stores the recently occurred virtual to physical address translations. But in a virtual environment these entries will be different for each guest OS. So, when the context switch between VMs happens, the TLB has to be flushed to avoid the guest OS accessing the translations of another guest OS. TLB flush on each context switch is computationally expensive \cite{Smith2006} \cite{tai2013comparisons}. Moreover, failure to hit the TLB causes many extra pipeline cycles.

  To overcome the problem, designers added an extra field named Address Space Identifier (ASID) to the TLB. The ASID field is used to distinguish the address of currently running process in a normal environment \cite{Smith2006}. The ASID field entry shows the owner (process) of the address translation entry in the TLB. In a virtualized system, a \textit{virtual TLB} is maintained by the hypervisor, which contains the virtual ASID field, virtual page field and the real page field. An \textit{ASID map table} is maintained by the hypervisor to map the \textit{virtual ASID}, of each process in each VM, to a unique \textit{real ASID} value. Only the address translations of the currently running process in the active guest OS, are allowed to access from the TLB (distinguished by the real ASID field) \cite{Smith2006}.

  Nested page tables along with either virtualized versions of architectured page tables or with virtualized versions of architectured TLBs, create a virtual address space in VMs. However, there are  issues with this addressing mechanism, from the perspective of performance and security. The nested paging technique compromises security while enabling Direct Memory Access (DMA) mechanism  \cite{Smith2006} \cite{DarrenAbramson2006} .

  The DMA mechanism is designed to work with the actual physical address space. If the DMA mechanism is enabled, the guest OS can reconfigure the device to access the memory of another VM through DMA mechanism \cite{szefer2012architectural}. A solution to this security threat, is to disable the DMA mechanism. Disabling DMA reduces performance because more CPU clock cycles would be required for data transfer between memory and I/O devices. Protecting memory by access from other VMs while enabling DMA mechanism in I/O devices, is called \textit{DMA isolation}. DMA isolation has to be achieved to enable the DMA mechanism and thereby improve the performance of VMs.

  The requirements for better performance without compromising security led to the development of another address translation mechanism named I/O Memory Management Unit (IOMMU) \cite{ben2006utilizing} which is explained next.

  \subsection{IOMMU}

  The IOMMU architecture is illustrated here with the help of an Intel based technology named Intel VT-d \cite{vt-directed-io-spec}. Intel introduced a DMA remapping hardware in their chip set. A generalized IOMMU architecture is implemented inside this DMA remapping hardware. The process of converting the DMA address from one form to another (virtual address to physical address) is known as DMA remapping. The hardware for DMA remapping is called DMA remapping hardware.

  Intel VT-d architecture divides the physical address space into different partitions. Each partition is a subset of the entire physical memory. Each partition is considered as a protection domain \cite{vt-directed-io-spec} \cite{DarrenAbramson2006}. I/O devices are allocated to a single protection domain by the hypervisor. I/O devices are not allowed to access any domain other than the one it is allocated. VMs are also allocated to a protection domain by the hypervisor. A VM is allowed to access only the I/O devices that are allocated to its own domain. VT-d enables hypervisor to allocate one or more I/O devices to a protection domain.

  DMA isolation is achieved by restricting the access to a protection domain by the I/O devices not assigned to that domain. DMA isolation is implemented through two address translation tables used in this architecture \cite{vt-directed-io-spec} \cite{DarrenAbramson2006}. They are 

  \begin{itemize}
   \item Root Entry Table (RET) 
   \item Context Entry Table (CET) 
  \end{itemize}

  VT-d hardware treats the address in a DMA request as a \emph{DMA Virtual Address(DVA)} \cite{vt-directed-io-spec} \cite{DarrenAbramson2006}. DVA can be a guest physical address (pseudo physical address) of the VM  to which the I/O device is assigned. Thus, I/O devices that are assigned to a protection domain can be provided a view of memory that is different from the host physical memory. VT-d transforms the DVA address to the host physical address with the help of RET and CET.

  Each DMA address has three fields \cite{vt-directed-io-spec} \cite{DarrenAbramson2006}. They are \emph{Bus number}, \emph{Device number} and \emph{Function number}. The bus number field content is   used to index into the RET. Each entry in the RET point to a CET. The device number and function number field contents are collectively used to index into the CET. Each entry in CET points to an address translation structure which is a multilevel page table similar to a IA-32 processor page table named \emph{hierarchical translation structure} \cite{vt-directed-io-spec}.

  Since VMs are in different protection domains, I/O devices alloted to a VMs protection domain would not be able to access the memory of another VM. Thus,  in VT-d architecture, the DMA mechanism can be enabled without compromising the security, thereby improving the performance of I/O devices, and consequently the performance of VMs.

  It is observed from the two memory management techniques, nested paging and IOMMU,  that nested paging  would have the  worse performance of the two,  due to the inability to safely use DMA. IOMMU architecture allows the use of DMA and could achieve DMA isolation, as the I/O devices and VMs are permitted to access only one protection domain. Memory translation is a two level lookup and is hence slower than a normal environment.

  Though the IOMMU architecture provides better DMA isolation than nested paging, it still stays vulnerable to many security threats. In order to illustrate the problem, we discuss the security threats and proposed solution in literature, in the succeeding paragraphs.

  \subsection{Security Threats}

  A survey on the security of VMs shows that there exist vulnerabilities at different architecture levels like CPU, memory and network, which would help a malicious VM  to easily gain the control of the hypervisor \cite{jithin2014virtual}. In the IOMMU architecture, the hypervisor has access to all the memory locations including the entire  allocated memory space for machines. This would result in the situation that a malicious VM infects a hypervisor or an infected hypervisor could attack or access the memory of another VM. Providing security in the presence of infected hypervisor is  considered  an open research problem in the area of the  security of VMs \cite{szefer2012architectural}. For improved security in the context of infected hypervisors, solutions that  move the protection of memory from hypervisor level to the hardware level are desirable \cite{szefer2012architectural}. The proposed work in \cite{szefer2012architectural}, called HyperWall, is 
briefly discussed.

  \subsubsection{HyperWall}

  \textit{HyperWall} \cite{szefer2012architectural} is an architectural solution  aimed at  providing hardware for protecting guest VMs from a malicious hypervisors. The HyperWall hardware is proposed as an extension to the IOMMU hardware unit. The authors  claim that the key feature of HyperWall is \emph{Confidentiality} and \emph{Integrity } protection to VMs. HyperWall architecture provides additional protection bits to each memory page without modifying the paging structure in the hypervisor or guest OS. Four protection bits are used by HyperWall associated with each physical page. These protection bits can represent four new protection modes to the physical page. They are \emph{Not assigned to any VM} (Hypervisor access only), \emph{Assigned to a VM with hypervisor and DMA allowed}, \emph{Assigned to a VM with hypervisor access denied} and \emph{Assigned to a VM with neither hypervisor nor DMA allowed.} The pages assigned to a VM are protected from other VMs and the hypervisor by applying these user 
specification protection modes to each page. 

  According to our analysis, there exist security issues with this architectural solution. They are

 \begin{itemize}

  \item HyperWall architecture aims to protect only the confidentiality and integrity of a VMs data and not \textit{availability}. However, availability is also an important security property. The architecture does not provide the assurance of  a minimum and fair amount of memory to all VMs at all point of their active lifetime. It allows the VM user to set the memory page with a protection level which denies access to DMA and hypervisor. Hence, any malicious VM can use this feature to create memory starvation for other VMs by locking several pages at a time.

  \item  HyperWall does not protect against covert channels and side channels attacks \cite{szefer2012architectural},which are the main threat towards the isolation property of VMs \cite{jithin2014virtual}. Most of the covert channel attacks are done through memory \cite{Xiao:2012:CCC:2382196.2382318}, \cite{barham2003xen} \cite{wu2011identification} \cite{Xu:2011:ELC:2046660.2046670}. Without preventing those attacks at the memory level, true isolation in the memory level cannot be achieved.

 \end{itemize}

  Apart from security threats, there are performance challenges to be met by virtualization mechanisms. In the next two paragraphs, we state  the major reasons for memory slowdown that result in under-performance, as identified in our survey.

 \subsection{Underutilization of Memory}

  Our survey on the challenges in memory architecture shows that a huge volume of unutilized memory is allocated to VMs \cite{agmon2014ginseng} \cite{hwang2014mortar}. It is because, in the current VM environment, the requested physical memory is divided and allocated to VMs when they are created. That memory would not be used by the assigned VMs if they do not need that memory. But, it cannot be assigned to other VMs that require memory. A fairer allocation of physical memory is required.

  \textit{Ginseng} \cite{agmon2014ginseng} is a solution to the problem of underutilization of VM memory. Ginseng runs in the hypervisor, and allocates physical memory to a guest OS via a balloon driver. The balloon driver is installed in each guest OS. A balloon controller is installed in hypervisor. The communication between balloon driver in guest OS and balloon controller in hypervisor is done through \textit{libvirt} \cite{agmon2014ginseng}, an application programming interface (API) for managing the hypervisor.

  There are two types of communication modes between the balloon driver and balloon controller, \textit{Inflation} and  \textit{Deflation}. In inflation, the balloon driver transfers memory from guest OS to the hypervisor. The hypervisor keeps such memory from each guest together as a memory pool, available to guest OS when its assigned memory is over-utilized, through deflation. By inflation and deflation, Ginseng solves the problem of memory starvation effectively.

  \textit{Mortar} \cite{hwang2014mortar} is another technique which helps to utilize this underutilized memory as a cache. There are two uses for this cache. The first one is to use it as a cache for pre-fetching disk blocks. The next is to use it as a an application level distributed cache that follows the \textit{memcached} \cite{wang2012streaming} protocol. Rather than allocating the underutilized memory to guest OS, this architecture pools together the spare memory on each guest OS and exposes it as a volatile cache. Whenever the guest OS require its memory, the memory cache objects are freed and given back to the guest OS. This architecture also efficiently uses the underutilized memory.

  The main difference between Ginseng and Mortar is that Ginseng pools the unallocated memory and gives it to other VMs, while Mortar pools the unallocated memory and uses it as an application level cache.
  
  Our analysis is that both Ginseng and Mortar are susceptible to covert channel based attacks by colluding VMs on the same server, thereby risking user program and data on the VMs.

  \subsection{Memory Access Times}
  Another major challenge in VM technology is to improve the access time of memory for processes. Both nested paging technique and IOMMU architecture use two level memory address translation. In nested paging architecture as well as in IOMMU architecture, address translation hardware is virtualized in such a way that the address translation hardware is directly accessible to hypervisor and not to guest OS \cite{Smith2006} \cite{DarrenAbramson2006}.  HyperWall architecture does not offer any improvement in VM performance over IOMMU or nested paging either.

  Existing literature also includes illustrations of challenges and solutions in deduplication \cite{chiang2013introspection} \cite{chen2014cmd} and double-paging \cite{arya2014tesseract} features of VMs. Deduplication and double-paging are implemented in hypervisors, by sharing the memory pages among VMs and hypervisor. Those features are considered as a threat to memory isolation \cite{jithin2014virtual}, and are hence not advisable for use in clouds.

  Enabling the access of address translation hardware directly by the guest OS removes one level of indirection in address translation process. This would improve the performance of VMs by improving the average memory access time. Clearly the need of the day is to have a memory virtualization infrastructure that supports the following properties even in the presence of a malicious hypervisors.
  \begin{itemize}
   \item Isolated physical memory region to each VM
   \item Guest OS should be able to use the address translation hardware directly
   \item Fair allocation of physical memory among VMs
  \end{itemize}

  \section{Architectural Support for Memory Isolation} \label{solution}

\begin{figure*}[ht!]
  \centering
  \includegraphics[width=100mm]{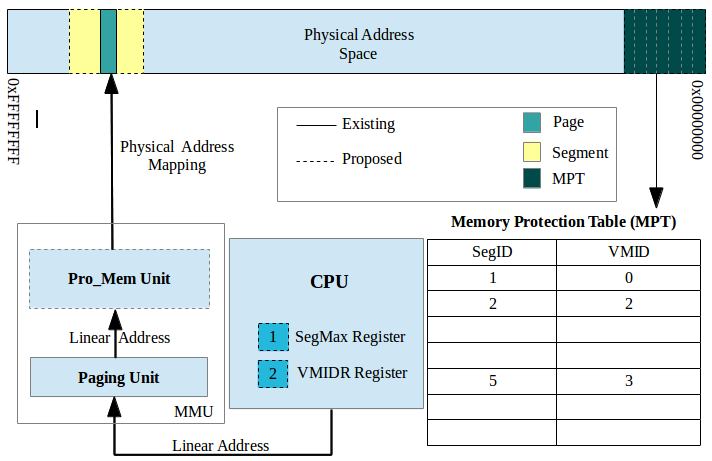}
  \caption{ASMI: Architectural Support for Memory Isolation}
  \label{proposal}
  \end{figure*}

   The analysis presented in the previous section clearly establishes that a novel mechanism has to be introduced, such that it would be able to protect the memory of a VM from a malicious hypervisor,  address the problem of underutilization of memory in VMs,  and allow the user programs to use the address translation hardware directly, for improved VM performance.   We propose  \textit{ASMI, Architectural Support for Memory Isolation},  that can satisfy the above requirements. Figure \ref{proposal} illustrates the proposed architecture.

   We propose ASMI as a generic solution, aimed at providing hardware for enabling direct access to physical memory by the VMs. We illustrate and expand the concept on an Intel platform with Intel-VT capabilities \cite{vt-directed-io-spec}, as details about the existing technology are available in literature.  Hence, hereafter, in this paper,  ASMI refers to the design enhancement we describe in the succeeding paragraphs.
   
   \paragraph{Description}In an Intel 64 bit architecture, the address translation mechanism disables  segmentation and only paging mechanism  exists \cite{64-ia-32-arch-manual-325462}. In ASMI,   segmentation is enabled and utilized to improve isolation. The physical memory is partitioned into physical segments. Each segment should be of a fixed length. Each segment contain fixed number of pages. A new hardware unit named \textit{Pro-mem} controls the entire physical memory.

  A new register named VMIDR is introduced and its contents managed by each processor to store the identity of  the currently running VM on that processor. A unique ID is assigned and stored in VMIDR register by the Pro-mem unit when a new VM is created. Switch between VMs on a processor causes a changes in the VMIDR value, done by the Pro-mem. Moving the control of execution from VM to hypervisor and vice versa would be informed to the Pro-mem by the \emph{VM Exit} and \emph{VM Entry} instructions.

  \emph{VM Entry} and \emph{VM Exit} are the instructions in Intel VT architecture \cite{vt-directed-io-spec} that move the execution to and fro between VMs and hypervisor. In the proposed architecture, these instructions are modified to inform the Pro-mem unit about the control transfer among VMs and hypervisor in an atomic manner.

  \emph{SegMax} is introduced, which is maintained by Pro-mem and contains the maximum number of segments(MSEG) that can be assigned to a VM when the entire physical memory is full.  The value stored in MSEG is calculated by dividing the total number of physical segments (TSEG) in primary memory with TOT, where \emph{TOT} is the sum of total number of VMs and hypervisor running above the hardware.

  SegMax, MSEG, and TOT values are not fixed. They are changed  when a new VM is created or when an old one is destroyed. TSEG is fixed at boot time, by the hypervisor, depending on the page size that the hypervisor is designed to work with. TSEG values cannot be changed until next reboot. Initially at boot time, MSEG and TOT values are zero. When a hypervisor is loaded or a VM is created, TOT value is incremented by one and MSEG is recalculated according to the new TOT value.

  A new data structure named \textit{Memory Protection Table (MPT)} is also proposed, and it is managed by the Pro-mem. MPT contains the segment ID (SegID) and its corresponding virtual Machine ID (VMID). Each segment can be assigned to a single VM.  MPT would be stored in the primary memory as shown in Figure \ref{proposal}. This primary memory portion of MPT will be inaccessible to any software module. This table is accessible only to the Pro-mem unit. A VMID value zero in MPT is used to indicate the segments that belongs to the hypervisor.

  Initially, when the physical machine boots, the MPT will be empty. When the hypervisor is started by the BIOS, Pro-mem stores a unique ID to the VMIDR register. Similarly when a VM is created, a unique ID is stored in VMIDR register by Pro-mem.

  When the \textit{VM exit} instruction is executed, VMIDR contents would get stored in the initial address of the first memory segment of the corresponding VM and the  VMIDR is loaded with the initial address of the first memory segment of the hypervisor. Similarly when the \textit{VM entry} instruction executes, the VMIDR value is  stored in the initial address of the first segment of hypervisor and the VMIDR value is loaded from the initial address of the first segment of the running VM. These load and store operations are executed by Pro-mem as atomic operations to \textit{VM exit} and \textit{VM entry} instructions.

  \paragraph{Allocation}
  When a VM requests for a memory page, Pro-mem would  check for available free pages in the allotted segments. If a free page is available in the allotted segment, it would  be assigned to the VM. If no free pages are available, Pro-mem would check for a free segment. If a free segment available, it is allotted to the current VM, the MPT updated and the page assigned to the VM.

  When no free segments are available, Pro-mem would  check whether any of the VMs has been allotted more than MSEG number of segments. The number of segments by which it exceeds  MSEG would be informed to the corresponding VM by the Pro-mem, to make it free through swapping. Then those pages would free be used by the requesting VM. If no VM is using more than MSEG segments, Pro-mem will send a memory full exception to the requesting VM. This exception can be used by the guest OS to swap its allotted memory pages and load the new one, and thereby continue its execution.

  The above techniques ensure a minimum number of physical segments or a minimum amount of physical memory to each VM when the physical memory need is the maximum. It simultaneously ensures that physical memory is not be left free or unutilized when a VM requires it. Hence, an optimum utilization of physical memory can be achieved by this architecture.

  \paragraph{Access}
  Access to  segments would  be validated by Pro-mem with the help of VMIDR value and MPT table. Pro-mem would raise an exception if a VM or hypervisor tries to access the segment which is not allotted to it.  The main advantage of this architecture is that it can provide memory isolation among VMs and hypervisor irrespective of the hypervisor security.

   In this version, we propose Pro-mem to be installed in between hardware paging architecture and primary memory. The paging unit would  be modified to include Pro-mem functionality. Each guest OS maintains a page table in memory, which contains the virtual address and its corresponding physical address. The guest OS gives the virtual/linear address to the paging unit. The paging unit would give this linear address to the Pro-mem unit. Pro-mem would return the physical address, within the guest OS alloted segment, to the paging unit. The paging unit would return this physical address to the guest OS.

  \paragraph{Performance and Security}
  The physical address obtained in guest OS, by the address translation described above, would be the actual physical address within the allotted physical segment. Only a single level address translation is required in this architecture. It would improve the performance of VMs, as it improves over two level translation. DMA can be enabled because the memory allotted to a VM can not be accessed by the other VMs as they are separated by different segments.  Separating the physical memory of each VMs at the hardware level could provide  hypervisor-independent security to VMs.

  Operating systems in a normal environment use the paging hardware to translate  linear addresses to  physical addresses. In a virtual environment, the guest operating system uses the paging hardware with Pro-mem to translate its linear address to the  physical address, and the physical address range is controlled by the Pro-mem. Thus, the guest operating system directly uses the address translation hardware (Paging unit with Pro-mem) similar to the normal (not virtualized) environment. Hence, no modification to guest OS is required. This enables the use of native operating systems, and a performance much better than existing virtualized systems.

  \section{Conclusions and Future Work} \label{conclusion}

    A memory architecture model named ASMI, has been proposed and described in the paper.   ASMI provides an isolated memory region to each VM and the hypervisor. ASMI has been illustrated in this paper on an Intel Platform with hardware enhancements to implement the design.  ASMI  is designed to provide memory isolation to VMs, irrespective of the integrity of hypervisor.

  Our architecture includes an address translation unit named Pro-mem. Pro-mem is shared among VMs in time division multiplexing without compromising security,  to improve address translation performance. The design of Pro-mem solves the problem of underutilization of memory, as Pro-mem allocates memory to VMs only on individual requests and does not pre-allocate. ASMI  can thus provide confidentiality, integrity and availability to VMs at memory level irrespective of hypervisor security without compromising performance.

  Implementing ASMI through kernel level simulations  and comparison of  the security and performance of VMs with the currently available memory architectures like IOMMU and nested paging are planned as future tasks  in our research. ASMI is a generic solution. Studying  the  suitability of ASMI  on other architectures like MIPS, and other RISC variants, is also included in our research agenda.
  
\bibliographystyle{IEEEtran}

\end{document}